\documentclass[sigconf]{acmart}
\usepackage{epsfig}
\usepackage{graphicx}
\usepackage{amsmath}
\usepackage{amssymb}
\usepackage{multirow}
\usepackage{bm}
\usepackage{hyperref}
\usepackage{balance}
\usepackage{color}

\AtBeginDocument{%
  \providecommand\BibTeX{{%
    \normalfont B\kern-0.5em{\scshape i\kern-0.25em b}\kern-0.8em\TeX}}}

\copyrightyear{2020} 
\acmYear{2020} 
\setcopyright{acmcopyright}
\acmConference[MM '20]{Proceedings of the 28th ACM International Conference on Multimedia}{October 12--16, 2020}{Seattle, WA, USA}
\acmBooktitle{Proceedings of the 28th ACM International Conference on Multimedia (MM '20), October 12--16, 2020, Seattle, WA, USA}
\acmPrice{15.00}
\acmDOI{10.1145/3394171.3413906}
\acmISBN{978-1-4503-7988-5/20/10}

\begin{document}
\fancyhead{}
\title{Adversarial Privacy-preserving Filter}

\author{Jiaming Zhang$^{1,2}$, Jitao Sang$^{1,2}$, Xian Zhao$^{1}$, Xiaowen Huang$^{1}$, Yanfeng Sun$^{3}$, Yongli Hu$^{3}$}
\affiliation{%
  \institution{$^{1}$School of Computer and Information Technology \& Beijing Key Lab of Traffic Data Analysis and Mining, Beijing Jiaotong University, Beijing, China}
}
\affiliation{%
  \institution{$^{2}$Peng Cheng Laboratory, ShenZhen, China}
}
\affiliation{%
  \institution{$^{3}$Beijing Key Laboratory of Multimedia and Intelligent Software Technology \& Beijing Artificial Intelligence Institute, Faculty of Information Technology, Beijing University of Technology, Beijing, China}
}

\email{lanzhang1107@gmail.com, {jtsang, 20120454, xwhuang}@bjtu.edu.cn, {yfsun,huyongli}@bjut.edu.cn}







\renewcommand{\shortauthors}{Jiaming Zhang, et al.}

\begin{abstract}

While widely adopted in practical applications, face recognition has been critically discussed regarding the malicious use of face images and the potential privacy problems, e.g., deceiving payment system and causing personal sabotage. Online photo sharing services unintentionally act as the main repository for malicious crawler and face recognition applications. This work aims to develop a privacy-preserving solution, called Adversarial Privacy-preserving Filter (APF), to protect the online shared face images from being maliciously used. We propose an end-cloud collaborated adversarial attack solution to satisfy requirements of privacy, utility and non-accessibility. Specifically, the solutions consist of three modules: (1) image-specific gradient generation, to extract image-specific gradient in the user end with a compressed probe model; (2) adversarial gradient transfer, to fine-tune the image-specific gradient in the server cloud; and (3) universal adversarial perturbation enhancement, to append image-independent perturbation to derive the final adversarial noise. Extensive experiments on three datasets validate the effectiveness and efficiency of the proposed solution. A prototype application is also released for further evaluation. We hope the end-cloud collaborated attack framework could shed light on addressing the issue of online multimedia sharing privacy-preserving issues from user side.\footnote{~\small{To encourage reproducible research, the code is available at {\color[RGB]{255,105,180}{\textbf{\href{https://github.com/adversarial-for-goodness/APF}{GitHub}}}}. Furthermore, a demo video illustrates the proposed APF solution in {\color[RGB]{255,105,180}{\textbf{\href{https://youtu.be/qO8d2LVECYk}{YouTube}}}}}.} 

\end{abstract}



\maketitle

\section{Introduction}\label{sec:1}

\begin{figure}[t]
  \begin{center}
    \includegraphics[width=\linewidth]{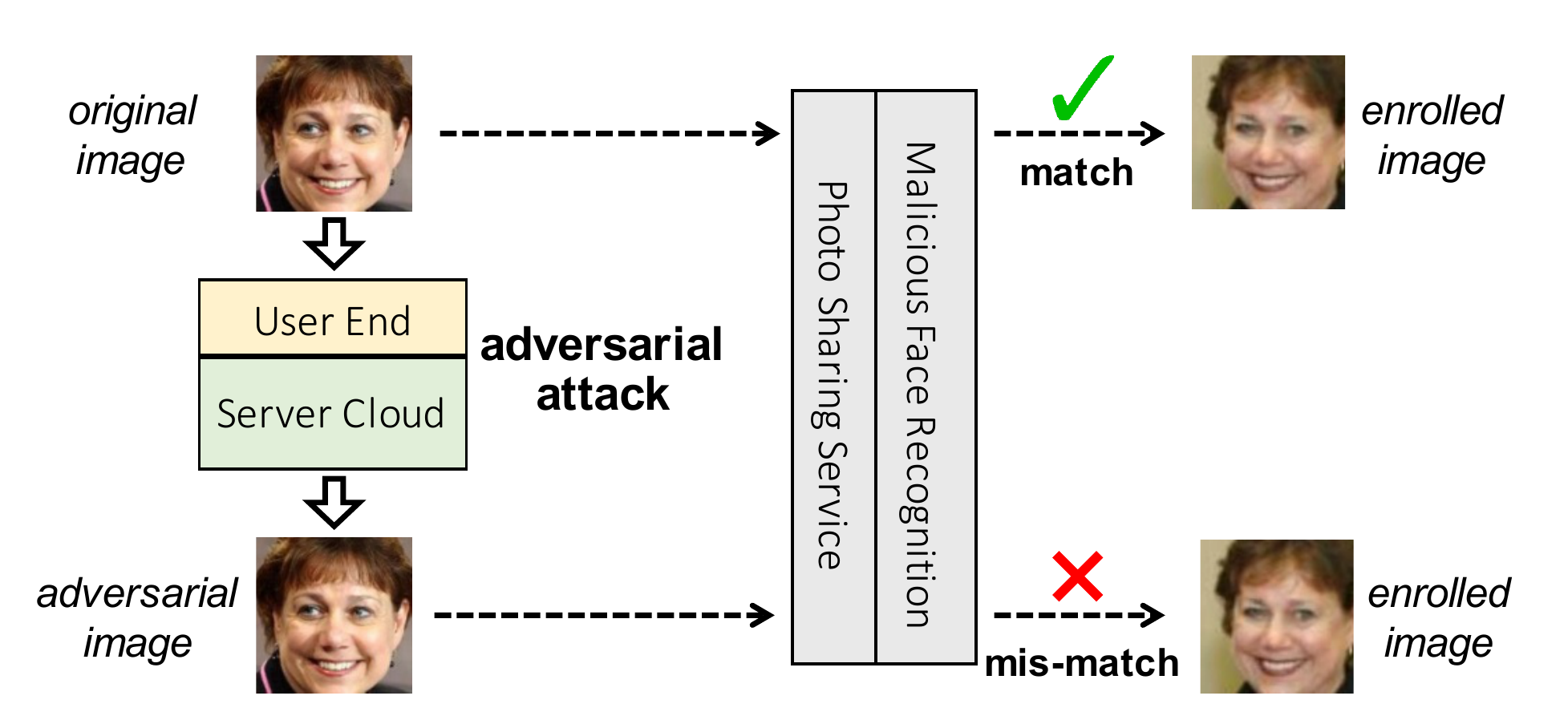}
  \end{center}
  \caption{Schematic illustration of the proposed adversarial privacy-preserving filter. Given a face image, the synthetic adversarial image is expected to fool the malicious face recognition algorithm.}
  \label{fig:1}
\end{figure}

Benefited from both algorithmic development and the adequate face image data, face recognition has been widely adopted in applications like criminal monitoring, security unlock, digital ticket and even face payment. However, the discussion of privacy problems caused by the unreasonable use of face images has never been stopped. Kneron tested that widely used face payment systems like AliPay and WeChat can be fooled by masks and face images, and Deepfake creating imitated videos causes severe personal sabotage~\cite{chesney2019deep}. Among the sources of personal face images, online photo sharing services unintentionally act as the main repository for malicious crawler and face recognition applications. Recently, a company called Clearview was disclosed to crawl face images from Facebook, YouTube and other websites, and construct a database containing more than three billion images. Therefore, it is critical to design privacy-preserving solutions for photo sharing services to protect the uploaded face images from being maliciously used.

This paper aims to preserve users' portrait privacy without affecting their photo sharing experience. Two requirements need to be first satisfied: (1) \emph{privacy}, that user's identification information is unrecognizable from the shared face images; (2) \emph{utility}, that the face image quality should not be undermined. It is interesting to note that adversarial attack~\cite{szegedy2014intriguing} exactly meets the above requirements: (1) the direct consequence of adversarial attack is to mislead model predictions which is consistent to fool the malicious face recognition algorithms; and (2) the introduced adversarial perturbation is trivial and imperceptible to human, thus retaining the image perception and utility for photo sharing. Therefore, we exploit adversarial attack to design the privacy-preserving filter, which adds adversarial perturbation to the original image before uploading to photo sharing services and expects to mislead the malicious face recognition algorithms (illustrated in Fig.~\ref{fig:1}).

While sharing the adversarial face image guarantees privacy-preserving on photo sharing services, there still remains risk of privacy leakage when generating the adversarial images. Since most adversarial attack solutions involve with complex models and massive computation operations, it is typical to upload the original face images to the privacy-preserving filter server for processing. This inevitably exposes the original face image to the third-party application and makes them vulnerable both during transmission and in the cloud. Therefore, a third requirement is needed for privacy-preserving filter: \emph{non-accessibility}, that the original face image should only be accessible in the user end. To address this, we propose an end-cloud collaborated adversarial attack solution, where the original face image is only used to extract image-specific gradient in the user end with a compressed probe model, and the image-specific gradient is then fine-tuned in the server cloud with state-of-the-art model to generate the final adversarial noise. The end-cloud collaborated solution succeeds to resolve the paradox between the computation shortage in the end and the privacy leakage risk in the cloud. Moreover, to improve the attack performance of the generated adversarial perturbation in the cloud, we further introduce the image-independent universal adversarial perturbation for enhancement, which demonstrates effectiveness in both accelerated training convergence and stronger attacking capability.    

The contributions of the paper can be summarized as follows:
\begin{itemize}
\item We designed an adversarial privacy-preserving filter to preserve users' portrait privacy from malicious face recognition crackers without affecting their photo sharing experience. Adversarial attack naturally meets the two basic requirements of \emph{privacy} and \emph{utility}.
\item We proposed an end-cloud collaborated adversarial attack framework, which addresses the additional \emph{non-accessibility} requirement to guarantee the original image only accessible to users' own device end. The compatible performance of this two-stage attack with the traditional one-stage attack (in Section~\ref{sec:4-2-3}) also provides a novel perspective to understand the adversarial attack problem.
\item We integrated the universal adversarial perturbation with image-dependent perturbation to obtain improved privacy-preserving capability. This sheds light on alternative way to exploit adversarial examples.
\item We conducted extensive experiments to validate the effectiveness and efficiency of the proposed solution framework. A prototype of adversarial privacy-preserving filter is further carried out and released for evaluation.
\end{itemize}

\section{Related Work}

\subsection{Privacy-preserving Photo Sharing}
With the increasing popularity of Online Social Networks(OSNs), privacy-preserving photo sharing has received considerable attention ~\cite{xu2016virtual,such2016resolving,li2018srim}. Existing attempts can be roughly categorized into preserving image metadata~\cite{zhang2016privacy}, setting access control protocols~\cite{klemperer2012tag} or sharing privacy policies~\cite{such2016resolving,sun2016processing}, and explicitly encrypting ROI region before uploading but decrypting after downloading~\cite{xu2016virtual,sun2016processing}. Different from these studies, the goal of this work is to hide the face identification information when sharing to OSN so that the potential image crawler and face cracker cannot use it for malicious usage.

\subsection{Adversarial Attack}

In the last few years, it has been witnessed that the existing machine learning models, not just deep neural networks, are vulnerable to adversarial attack~\cite{szegedy2014intriguing}. ~\citet{szegedy2014intriguing} first introduced the problem of adversarial attack, and proposed a box-constrained L-BFGS method to find adversarial examples. To address the expensive computation, ~\citet{goodfellow2014explaining} proposed Fast Gradient Sign Method (FGSM) to generate adversarial examples by performing a single gradient step, which later becomes a widely-used baseline attack method. ~\citet{kurakin2016adversarial} extended this method to an iterative version (I-FGSM). ~\citet{dong2018boosting} further improved adversarial examples by adopting momentum term (MI-FGSM). ~\citet{xie2019improving} proposed D$\rm I^2$-FGSM and M-D$\rm I^2$-FGSM by adopting input diversity strategy, which focused on improving the transferability and achieved state-of-the-art performance in black-box setting. 

There exist many attempts to explore the adversarial attack problem in face recognition applications. ~\citet{sharif2016accessorize} and ~\citet{komkov2019advhat} proposed face recognition attacks by modifying facial attributes like adding virtual eyeglasses to impersonate other subjects or prevent them from being recognized. ~\citet{dong2019efficient} proposed a query-based method for generating adversarial faces in black-box setting, which requires at least $1,000$ queries to the face recognition system. Some other works, e.g., AdvFaces~\cite{deb2019advfaces} focused on employing Generative Adversarial Network (GAN)~\cite{goodfellow2014generative} to craft new adversarial images. However, although the above face recognition attack solutions easily satisfy the \emph{privacy} requirement mentioned in Section~\ref{sec:1}, they may violate either the \emph{utility} or \emph{non-accessibility} requirement: (1) Attack solutions like modifying facial attributes and employing GAN tend to introduce non-trivial perturbation and make the perturbed face images unsuitable for sharing. (2) The above solutions all request to access the original face images, which leaves privacy-leakage risk in the cloud. Different from the existing face recognition attack solutions, in this work, we propose an end-cloud collaborated adversarial attack solution to simultaneously satisfy the \emph{privacy, utility} and \emph{non-accessibility} requirements. Moreover, the solution is compatible to allow instantiation with most of the existing adversarial attack algorithms.

In addition to the above image-dependent adversarial attacks, Moosavi-Dezfooli et al.~\cite{moosavi2017universal} found a type of image-independent noise called universal adversarial perturbation (UAP), which can mislead the pre-trained model for different images. Following this, Poursaeed et al.~\cite{poursaeed2018generative} trained a generative network for generating universal adversarial perturbations (GAP), and ~\citet{li2019regional} observed the property of regional homogeneity and generated regionally homogeneous perturbations (RHP). These studies demonstrate the possibility of attacking the classification of specific images with universal perturbation. Inspired by this, on the basis of image-specific perturbations, we further enhance the adversarial examples with universal adversarial perturbations in the cloud, with improved training convergence as well as attack performance verified in the later experiments.

\begin{figure*}[ht]
  \centering
  \includegraphics[width=\linewidth]{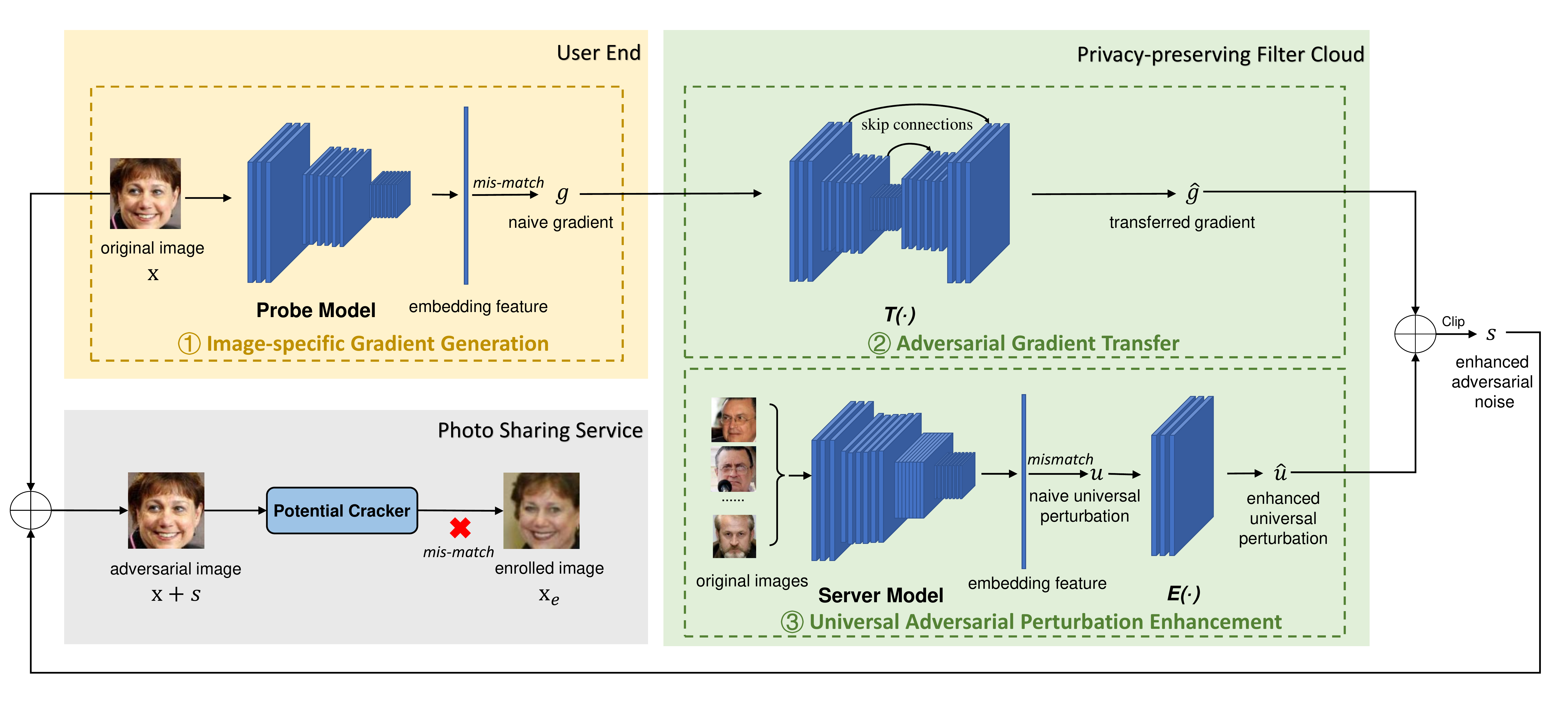}
  \caption{The proposed adversarial privacy-preserving filter framework.}
  \label{fig:framework}
\end{figure*}

\section{METHODOLOGY}

\subsection{Problem Definition and Notations}

\textsc{Definition 1} (\textsc{Adversarial Privacy-preserving Filter}). 

Given an original image $\mathbf{x}$, we assume that a face recognition model $f_\theta$ outputs an $l$-dimension vector as embedding feature $f_\theta(\mathbf{x})$. $\mathbf{x_e}$ represents the corresponding enrolled image, when the distance $d(f_\theta(x), f_\theta(x_e))$ between the two images is less than a certain threshold, they are judged to be the same subject, otherwise they are different subjects. 

The main focus of this paper is to design the \emph{adversarial privacy-preserving filter} (APF) that is an adversarial invisible noise $s$, making the adversarial example $\mathbf{x}+s$ look the same with $\mathbf{x}$ from the perspective of human observation but does not belong to the same subject from the perspective of face recognition model $f_\theta$. Under the condition of ensuring image quality, it preserves users’ portrait privacy without affecting their photo sharing experience.

\begin{table}[t]
\centering
\caption{Notation and explanations.}\label{tab:notation}
\begin{tabular}{l|l} 
\toprule[1pt]
\textbf{Symbol} & \textbf{Notation}\\
\midrule
\midrule
$\mathbf{x}$ & original image \\
\hline
$\mathbf{x_e}$ & enrolled image \\
\hline
$f_\theta$ & face recognition model with parameter $\theta$ \\
\hline
$g$ & naive gradient generated on probe model \\
\hline
$\tilde{g}$ & gradient generated on server model \\
\hline
$\hat{g}$ & transferred gradient \\
\hline
$u$ & universal perturbation \\
\hline
$\hat{u}$ & enhanced universal perturbation \\
\hline
$s$ & enhanced adversarial noise \\
\hline
$\epsilon$ & maximum change restriction of each pixel  \\
\hline
$d$ & distance function \\
\hline
$L$ & loss function \\
\hline
$T$ & gradient transfer module  \\
\hline
$E$ & enhancement module  \\
\hline
$APF\_g$ & adversarial image based on naive gradient \\
\hline
$APF\_\hat{g}$ & adversarial image based on transferred gradient \\
\hline
$APF\_s$ & adversarial image based on enhanced adversarial noise \\
\hline
$Images\_u$ & adversarial image based on universal perturbation \\

\bottomrule[1pt]
\end{tabular}
\end{table}

The main notations of this work are summarized in Table~\ref{tab:notation}.

\subsection{Overall Framework}
The working flow of privacy-preserving filter is illustrated in Fig.~\ref{fig:framework}, where three parts are involved as user end, privacy-preserving server cloud and photo sharing service. Given a portrait image, its image-specific gradient is first extracted in the user end, and then issued to the privacy-preserving server cloud for enhancement to derive adversarial noise. After perturbed by the enhanced adversarial noise, the resultant adversarial portrait image is expected to fool the potential face cracker on photo sharing service. It is noted that the original image is accessible only to user end during the whole process, preventing information leakage even on the privacy-preserving server.

The core solution is an end-cloud collaborated adversarial attack framework, consisting of three algorithm modules deployed respectively in the user end and server cloud: (1) \emph{Image-specific gradient generation} is to obtain the image-specific gradient by employing a compressed probe model runnable in the end. (2) \emph{Adversarial gradient transfer} module is to align the image gradient from probe model and the server model in the cloud, with the goal to recover the adversarial information fooling practical crackers. (3) \emph{Universal adversarial perturbation enhancement} module is to append image-independent universal perturbation to further enhance the derived adversarial noise. The following will elaborate each module.

\subsection{Image-specific Gradient Generation}\label{sec:3-1}

Following the \emph{non-accessibility} requirement, users' original face images should not be uploaded to the cloud server to avoid leakage. However, most of the existing adversarial attack algorithms are too complicated to be deployed in the end due to their high demands on computing resources. Therefore, our solution is to employ a compressed model in the user end to extract preliminary information from original image to mismatch the enrolled image, and then enhance the information in the cloud. 

Specifically, we use the compressed model as probe model in the user end to extract image-specific adversarial gradient $g$. Noted that different adversarial attack algorithms are allowed, which lead to $g$ with different levels of intensity under the same maximum perturbation. In experiments we will discuss the influence of different adversarial gradient extraction algorithms. In this subsection, we employ FGSM as the example algorithm to introduce this process. Following standard adversarial attack operation, image-specific gradient $g$ for original image $\mathbf{x}$ is extracted by:
\begin{equation}\label{eq1}
  g = \epsilon \cdot sign(\nabla_{\mathbf{x}} L(\mathbf{x}, \mathbf{x_{e}}; \theta))
\end{equation}
where $\mathbf{x_{e}}$ is the corresponding enrolled image, $\theta$ denotes the network parameters, $\epsilon$ ensures that the generated gradient is within the $\epsilon$-ball in the $L_\infty$ space, and sign($\cdot$) denotes the sign function. $L(\cdot)$ is the loss function that measures the distance between the original face image and the enrolled image:
\begin{equation}\label{eq2}
  L(\mathbf{x}, \mathbf{x_{e}}; \theta) = -d (f_\theta(\mathbf{x}), f_\theta(\mathbf{x_{e}}))
\end{equation}
where $f_{\theta}(\cdot)$ represents the embedding feature, $d(\cdot,\cdot)$ is distance function. We use Euclidian distance in our work.

\subsection{Adversarial Gradient Transfer}\label{sec:3-2}

The small-scale probe model extracts the image gradient in the user end for generating adversarial samples. However, since the structure and parameters of probe model in the end and server model in the cloud are different, the gradient cannot be used directly. To align the image gradient between probe model and server model, we propose the gradient transfer module $T$, which is defined as:
\begin{equation}\label{eq4}
  \hat{g} = T(g)
\end{equation}

We consider the gradient transfer module as an image-to-image translation network. The U-Net architecture is widely used in image-to-image translation problem, which is an encoder-decoder network with skip connections between the encoder and the decoder~\cite{ronneberger2015u}. We employ this architecture for transferring gradients from probe model to server model. Specifically, we train the gradient transfer module as the following optimization problem:
\begin{equation}\label{eq5}
  \min_{\theta_T} \Vert \hat{g} - \tilde{g} \Vert_2
\end{equation}
where $\tilde{g}$ is the gradient generated on server model following Eqn.~\eqref{eq1} and $\theta_T$ represents the parameters of network $T$. This objective function enforces the network $T$ to learn the correlation between the probe model and server model.

Potential crackers may use multiple face recognition models, which is essentially a black-box adversarial attack problem. It is recognized that adversarial perturbation is transferable between models: if an adversarial image remains effective for multiple models, it is more likely to transfer to other models as well~\cite{liu2017delving}. Inspired by this, to improve the attack performance to unknown cracking models, we implement the gradient transfer module not only for one-to-one domain transfer but also for one-to-many.

Specifically, given $K$ white-box face recognition models with their corresponding gradient $\tilde{g}_1$, ..., $\tilde{g}_k$, we re-formulate the loss function in Eqn.~\eqref{eq5} by replacing $\tilde{g}$ with $\tilde{g}_{ensemble}$:
\begin{equation}\label{eq6}
  \tilde{g}_{ensemble} = \sum_{k=1}^{K} \alpha_k \tilde{g}_k
\end{equation}
where $\alpha_k$ is the ensemble weight with $\sum_{k=1}^{K} \alpha_k = 1$. For many face recognition models, the larger the value of $K$, the stronger the generalization capability of the derived adversarial images. However, an excessive $K$ value will lead to high computational complexity and trivial weight $\alpha_k$ to underemphasize single model. We select $K = 2$ and evenly set $\alpha_k=1/2$~\footnote{~\small{Usually $\alpha_k=1/K$ except when prior is available to emphasize on some of the models.}}. In this study, we respectively select two of three state-of-the-art face recognition models as ensemble models for training and the remaining one as black-box model for testing. The performance of employing ensemble adversarial training to resist different face recognition models is reported in Section~\ref{sec:4-3-1}.

\subsection{Universal Adversarial Perturbation Enhancement}

As claimed in previous studies~\cite{moosavi2017universal,poursaeed2018generative,li2019regional}, universal adversarial perturbation contains image-independent information to mislead the classification of multiple images.  This inspires us to integrate the image-specific information and image-independent information to enhance the performance of adversarial perturbation. Referring to the previous study~\cite{moosavi2017universal}, universal adversarial perturbation is a vector $u$ that causes label change for most images sampled from the data distribution $\eta$:
\begin{equation}\label{eq7}
  \hat{C}(\mathbf{x}+u) \neq \hat{C}(\mathbf{x}) \quad \rm{\; for \;most }\; \mathbf{x} \sim \eta
\end{equation}
where $\hat{C}$ is a classification function that outputs for each image $\mathbf{x}$ an estimated label $\hat{C}(\mathbf{x})$. The previous universal adversarial perturbation studies produce a universal perturbation with the goal to cause image misclassification~\cite{moosavi2017universal,poursaeed2018generative,li2019regional}. Instead of directly misleading image classification, we expect such a universal adversarial perturbation $u$, that provides a fixed image-independent perturbation deviating the examined face image far away from other face images. Therefore, we derive the naive universal perturbation as follow:
\begin{equation}\label{eq8}
  u = \max_u \sum_{i=1}^n \sum_{j=1}^n d(f_\theta(\mathbf{x}_i+u),f_\theta(\mathbf{x}_j))
\end{equation}
where $n$ represents the number of all images contained in the dataset. 

To integrate the image-specific perturbation and image-indepen-dent perturbation, we further design an enhancement module to adapt the domain of the universal perturbation $u$ with the domain of $\hat{g}$. The enhancement module $E$ consists of a scale transformation layer and a $1\times 1$ convolution layer, which is defined as follows:
\begin{equation}\label{eq9}
  \hat{u} = E(u) = {\rm{conv}} (\beta \cdot u + \gamma)
\end{equation}
where $\beta$ and $\gamma$ are trainable parameters and $\rm{conv(\cdot)}$ is the $1\times 1$ convolution layer. The overall optimization problem incorporating the two proposed modules is as follows:
\begin{equation}\label{eq10}
  \max_{\theta_T, \theta_E} \;d(f_\theta(\mathbf{x}+ {\rm{clip}}_{\epsilon}  (\hat{g}+\hat{u})),f_\theta(\mathbf{x}_e))
\end{equation}
where ${\rm{clip}}_{\epsilon}(\cdot)$ indicates clipping the input within the $\epsilon$-ball, $\theta_T$ denotes the parameters of network $T$, and $\theta_E$ denotes the parameters of network $E$.

Specifically, to employ ensemble adversarial attack, similar to Eqn.~\eqref{eq6}, given $K$ face recognition models with their corresponding embedding feature $f_1,...,f_k$, we reformulate Eqn.~\eqref{eq10} to derive the final objective function by replacing $f_\theta(\cdot)$ with $\tilde{f}(\cdot)$ defined as follows:
\begin{equation}\label{eq:insert}
  \tilde{f}(\cdot) = \sum_{k=1}^{K} \alpha_k f_k(\cdot)
\end{equation}
Network $\emph{E}$ and $\emph{T}$ are then trained by solving the above final objective function. The output $\hat{u}$ and $\hat{g}$ of $\emph{E}$ and $\emph{T}$ are added to derive $s$. The cloud will output $s$, and then return to user end to add it to the original image to derive the privacy preserved image, which can be safely shared to OSNs.

\section{EXPERIMENTS}\label{sec:4}

\subsection{Experimental Settings}\label{sec:4-1}
\paragraph{{\rm{\textbf{Datasets}}}}
We train the gradient transfer module $T$ and the enhancement module $E$ on the subset of MS-Celeb-1M dataset~\cite{guo2016ms}. LFW dataset~\cite{huang2008labeled}, AgeDB-30 dataset~\cite{moschoglou2017agedb} and CFP-FP dataset~\cite{sengupta2016frontal} are used as test datasets to verify the effectiveness of the proposed privacy-preserving filter. We aim to jam the malicious face recognition algorithm to match the same users, thus the positive pairs (belong to the same person) are used for testing. In the experiments, we select subjects with two face images, where one is used as the enrolled image, and the other is for synthesizing the adversarial image.

\begin{itemize}
    \item \textbf{MS-Celeb-1M} contains $10$ million images. We randomly select $50,000$ face images of $1,000$ subjects for training, where each subject contains $50$ face images.
    \item \textbf{LFW} contains $13,233$ images of $5,749$ different subjects. According to the refined version of \citet{deng2019arcface}, we use $6,000$ images to construct $3,000$ positive pairs of images. 
    \item \textbf{AgeDB-30} contains $16,488$ images of $568$ different subjects. Same as above, we use $6,000$ images to construct $3,000$ positive pairs of images. 
    \item \textbf{CFP-FP} contains $7,000$ images of $500$ different subjects. Same as above,  we use $7,000$ images to construct $3,500$ positive pairs of images.
\end{itemize}

\paragraph{{\rm{\textbf{Face Recognition Models}}}}

In this study, we employ $4$ state-of-the-art face recognition models to verify the effectiveness of the adversarial examples generated by our privacy-preserving models. MobileFaceNet~\cite{chen2018mobilefacenets} is served as probe model, and ArcFace~\cite{deng2019arcface} is used as default server model. To demonstrate the resistance to unknown cracking models, we introduce ArcFace, FaceNet~\cite{schroff2015facenet} and SphereFace~\cite{liu2017sphereface} as server models in Section~\ref{sec:4-3-1}.

\begin{itemize}
    \item \textbf{MobileFaceNet} is specifically tailored for high-accuracy real-time face verification on mobile devices with $5.20$MB model size, whose backbone network is MobileNet-V2~\cite{sandler2018mobilenetv2}. It was trained on MS-Celeb-1M dataset.
    \item \textbf{ArcFace} is the best public Face ID system, whose backbone network is Resnet-v2-152~\cite{he2016identity}. It was trained on MS-Celeb-1M dataset.
    \item \textbf{FaceNet} and \textbf{SphereFace} were trained on CASIA-WebFace dataset~\cite{yi2014learning}. The backbone networks of FaceNet and Sphere-Face are Inception-Resnet-v1~\cite{szegedy2017inception} and Sphere20~\cite{liu2017sphereface}, respectively.
\end{itemize}

\paragraph{{\rm{\textbf{Evaluation Metrics}}}}

Recalling that the positioned privacy-pres-erving filtering problem is to both invalid the potential face recognition cracker and retain the original image quality, we introduce evaluation metrics regarding these two goals respectively. 

For invaliding crackers, we use the standard attack success rate (ASR)~\cite{xie2019improving} as the evaluation metric.

\begin{itemize}
    \item \textbf{ASR} measures the effectiveness of our adversarial perturbation: 
    
    \begin{equation}\label{eq11}
        {\rm{ASR}} = \frac{N_{w/o} - N_{w/}}{N_{total}}
    \end{equation}
    where $N_{w/o}$ and $N_{w/}$ denote the number of correctly recognized face images without and with perturbation\footnote{~\small{In this study, we use \textbf{$\bm{L_\infty}$ distance} to restrict each pixel's change within a maximum scale, but with no limit on the number of pixels that are modified.}}, respectively. $N_{total}$ denotes the total number of face images. The higher the ASR value, the better the adversarial perturbation effect, and the more satisfied the \emph{privacy} requirement.
    
\end{itemize}    
For retaining the original image quality, we quantify the similarity between the perturbed images and the original images via structural similarity (SSIM)~\cite{wang2004image}.
\begin{itemize}
    
    \item \textbf{SSIM} is a normalized metric whose values range from $-1$ to $1$, which means the similarity from completely different image pairs to identical image pairs:
    
        \begin{equation}\label{eq13}
        SSIM(x,y)=\frac{(2\mu_x\mu_y+c_1)(2\sigma_{xy}+c_2)}{(\mu_x^2+\mu_y^2+c_1)(\sigma_x^2+\sigma_y^2+c_2)}
    \end{equation}
    
    Here, $x$ and $y$ are the two images to be compared, $\mu_x,\, \mu_y,\, $ $\sigma_x, \sigma_y,\, \sigma_{xy}$ are the means and variances of $x$ and $y$, the covariance of $x$ and $y$, respectively. $c_i = (d_iJ)^2$, where $J=(2^{({\rm \sharp \ of \ bits \ per \ pixel})}-1), \,  d_1=0.01, \,d_2=0.03$ by default~\cite{wang2003multiscale}.
    
\end{itemize}

\begin{table*}[t]
\centering
\caption{The ASRs on LFW, AgeDB-30 and CFP-FP datasets.}\label{tab:1}
\begin{tabular}{c|c|ccccc||ccc} 
\toprule[1pt]
Datasets & Adversarial Images & FGSM & I-FGSM & MI-FGSM & DI$^2$-FGSM & M-DI$^2$-FGSM & UAP & GAP & RHP\\
\midrule
\multirow{5}*{LFW}
    & $APF\_g$        & 74.5\%	& 86.8\% & 88.2\% &	92.4\% & 89.1\% & - & -& -\\
    & $APF\_{\hat{g}}$& 91.9\%	& 95.4\% & 89.8\% &	96.9\% & 96.5\% & -      & -     \\
    & $APF\_s$      & 94.8\%	& 97.4\% & 95.7\% &	\textbf{98.8\%} & \textbf{98.8\%}&-&-\\
    & $Images\_u$      & - & -& - & -& - &27.9\% &11.3 &4.3\%\\
\cline{2-10}    
    & original image accessible      & 98.5\% & 99.4\% & 99.4\% &99.4\% & 99.27\%&-&-& -\\
\midrule
\midrule
\multirow{5}*{AgeDB-30}
    & $APF\_g$        & 81.7\% & 86.3\% & 88.1\% & 90.5\% &	88.6\% & - & - & -\\
    & $APF\_{\hat{g}}$& 82.3\% & 90.8\% & 90.8\% & 94.9\% &	92.8\% & -      & -  & -   \\
    & $APF\_s$      & 88.3\% & 93.4\% & 93.8\% & \textbf{95.5\%} &	94.7\%&-&-& -\\
    & $Images\_u$      & - & -& - & -& - &22.0\% &23.7 &13.1\%\\
\cline{2-10}
    & original image accessible      & 95.8\% & 96.0\% & 96.0\% & 96.0\% &	96.0\%&-&-& -\\
\midrule
\midrule                       
\multirow{5}*{CFP-FP}
    & $APF\_g$        & 48.6\% & 57.2\% & 63.8\% & 68.3\% &	65.0\% & - & -& -\\
    & $APF\_{\hat{g}}$& 51.8\% & 72.8\% & 74.9\% & 84.7\% &	78.1\% & -      & -  & -   \\
    & $APF\_s$      & 67.4\% & 79.6\% & 82.8\% & \textbf{88.3\%} &	85.3\%&-&-& -\\
    & $Images\_u$      & - & -& - & -& - &8.0\% &6.1\% &3.1\%\\
\cline{2-10}
    & original image accessible      & 92.5\% & 93.7\% & 90.8\% & 93.2\% &	93.4\%&-&-& -\\
\bottomrule[1pt]
\end{tabular}
\end{table*}

\paragraph{{\rm{\textbf{Implementation Details}}}}

We adopt ADAM optimizer with a fixed learning rate of $0.0001$ for the entire network. Each mini-batch consists of $100$ face images. The maximum perturbation $\epsilon$ of each pixel is set to be $8$, which is more imperceptible for human observers than general adversarial perturbation~\cite{luo2015foveation}. The embedding feature dimension $l$ of MobileFaceNet, ArcFace, FaceNet and SphereFace is set to $192$, $512$, $128$ and $512$, respectively.  Different face recognition models have different input sizes. We resize the input images to $112\times 112\times 3$ for ArcFace and MobileFaceNet, $160 \times160\times 3$ for FaceNet and $96\times 96\times 3$ for SphereFace. For adversarial gradient extraction algorithms (I-FGSM, MI-FGSM, DI$^2$-FGSM and M-DI$^2$-FGSM) are as follows: (1) For stochastic image transformations, we consider $3$ transformations: rescaling, rotation and color conversion. (2) The probability of transformations is set to be $0.5$. (3) The step size is set to be $2$. (4) The total iteration number is set to be $30$. (5) The decay factor of momentum term is set to be $1$.

\begin{figure}[t]
  \begin{center}
    \includegraphics[width=\linewidth]{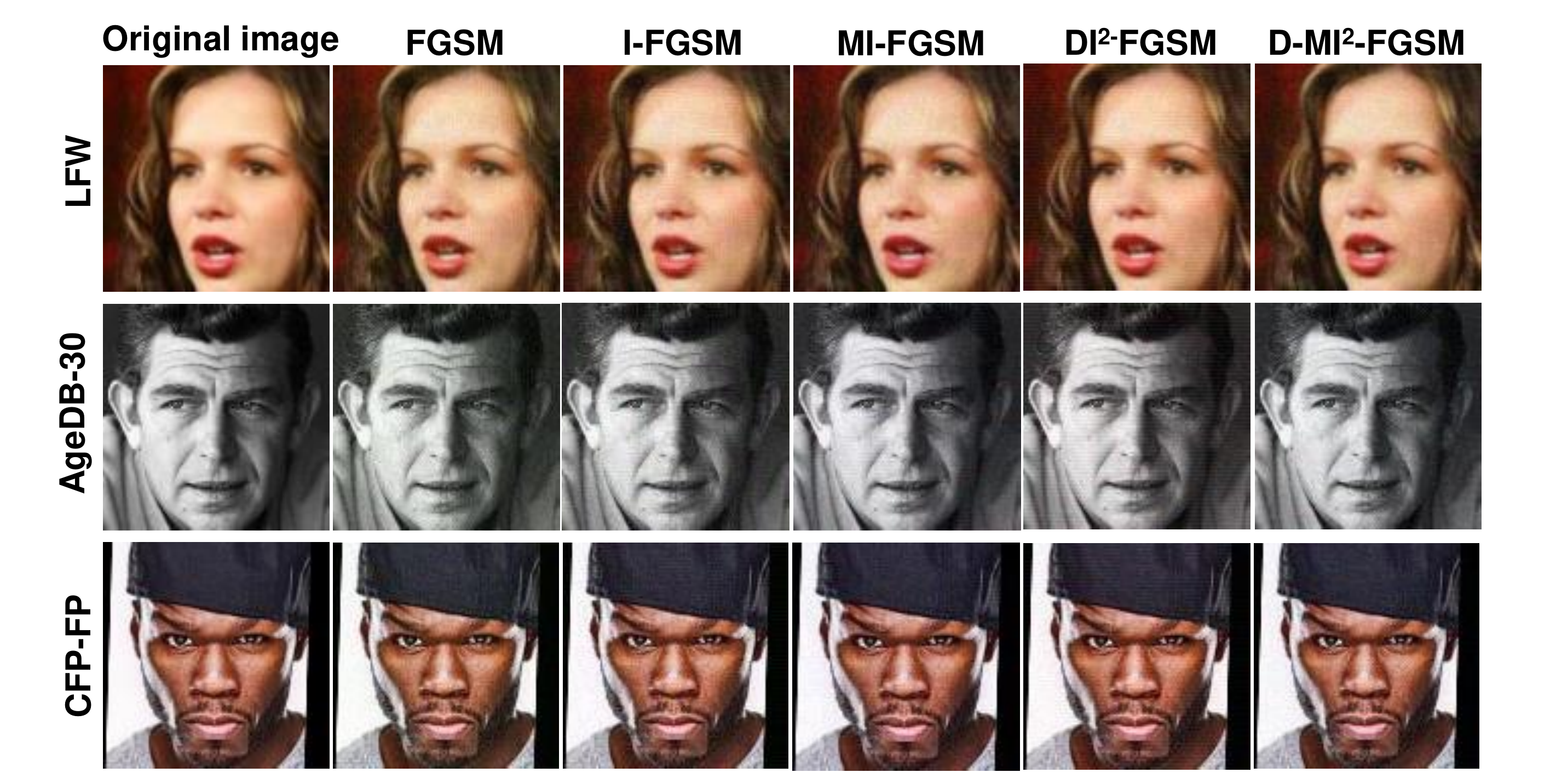}
  \end{center}
  \caption{The example adversarial images perturbed by $s$. Three rows correspond to three different datasets, and each column corresponds to a different gradient extraction algorithm in Table~\ref{tab:1}.}
  \label{fig:3}
\end{figure}

\begin{figure}[t]
  \begin{center}
    \includegraphics[width=\linewidth]{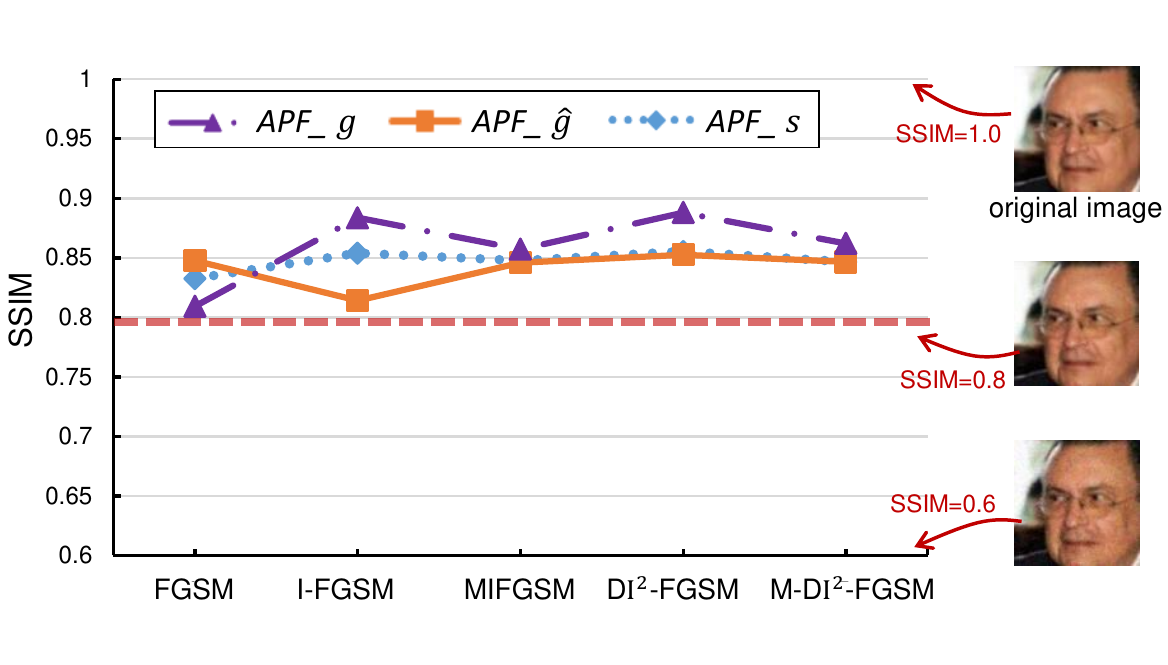}
  \end{center}
  \caption{The SSIMs between the generated adversarial images and original images. The reference line $0.8$ indicates the high quality of the adversarial images.}
  \label{fig:4}
\end{figure}


\subsection{Performance Evaluation}\label{sec:4-2}
\subsubsection{On \textbf{Privacy} and \textbf{Utility}}\label{sec:4-2-1}

Traditional adversarial attack methods are usually under the assumption of being accessible to original images. In this subsection, on the premise of satisfying the \emph{non-accessibility} requirement, we discuss the experimental results of our end-cloud collaborated solution on different adversarial gradient extraction algorithms and datasets. 

To demonstrate how our method meets the \emph{privacy} requirement, the ASRs of all the compared methods based on $5$ adversarial gradient extraction algorithms (FGSM, I-FGSM, MI-FGSM, DI$^2$-FGSM and M-DI$^2$-FGSM) are examined and shown in Table~\ref{tab:1}. $APF\_g$, $APF\_{\hat{g}}$ and $APF\_s$ indicate the output adversarial images at three processing stages: generating images only based on the naive adversarial gradient $g$ extracted by probe model in the user end, generating images based on the transferred gradient $\hat{g}$ in the cloud, and generating images based on the enhanced adversarial noise $s$ which contains image-specific noise and image-independent universal perturbation. We have the following main findings: (1) In the case of meeting \emph{non-accessibility} requirement, the ASRs of our proposed method on the three datasets reach $98.8\%$, $95.5\%$ and $88.3\%$, respectively. The good attack performance proves that the proposed filter can preserve user's \emph{privacy} by fooling face recognition models. (2) Our privacy-preserving filter is compatible with various gradient extraction algorithms. The stronger the gradient extraction algorithms (viewing the results of the five algorithms shown in Table~\ref{tab:1} from left to right), the better performance the filter achieves. This suggests us to use more effective adversarial gradient extraction algorithm when deploying the proposed framework in practical applications. (3) For each gradient extraction algorithm, $APF\_{\hat{g}}$ performs better than $APF\_g$ and $APF\_s$ achieves the best performance, which verify the effectiveness of the gradient transfer module and universal adversarial perturbation module.


To demonstrate how our method meets the \emph{utility} requirement, some example images corresponding to $APF\_s$ mentioned in Table~\ref{tab:1} are shown in Fig.~\ref{fig:3}. We can see that it is hard to tell the difference between original images and synthesized adversarial images from the perspective of human observation. We further use the quantitative metrics to measure image quality. SSIMs between the processed images and the original images on LFW dataset are illustrated in Fig.~\ref{fig:4}. We can observe that, when SSIM value of an image is greater than $0.6$, the human eye can hardly tell the difference. The SSIMs of our processed images are all greater than $0.8$, which indicates that the adversarial examples generated by our method can not only effectively interfere with the crack of face recognition, but also provide users with high-quality images to upload to the OSNs for use.

\begin{figure}[t]
  \begin{center}
    \includegraphics[width=\linewidth]{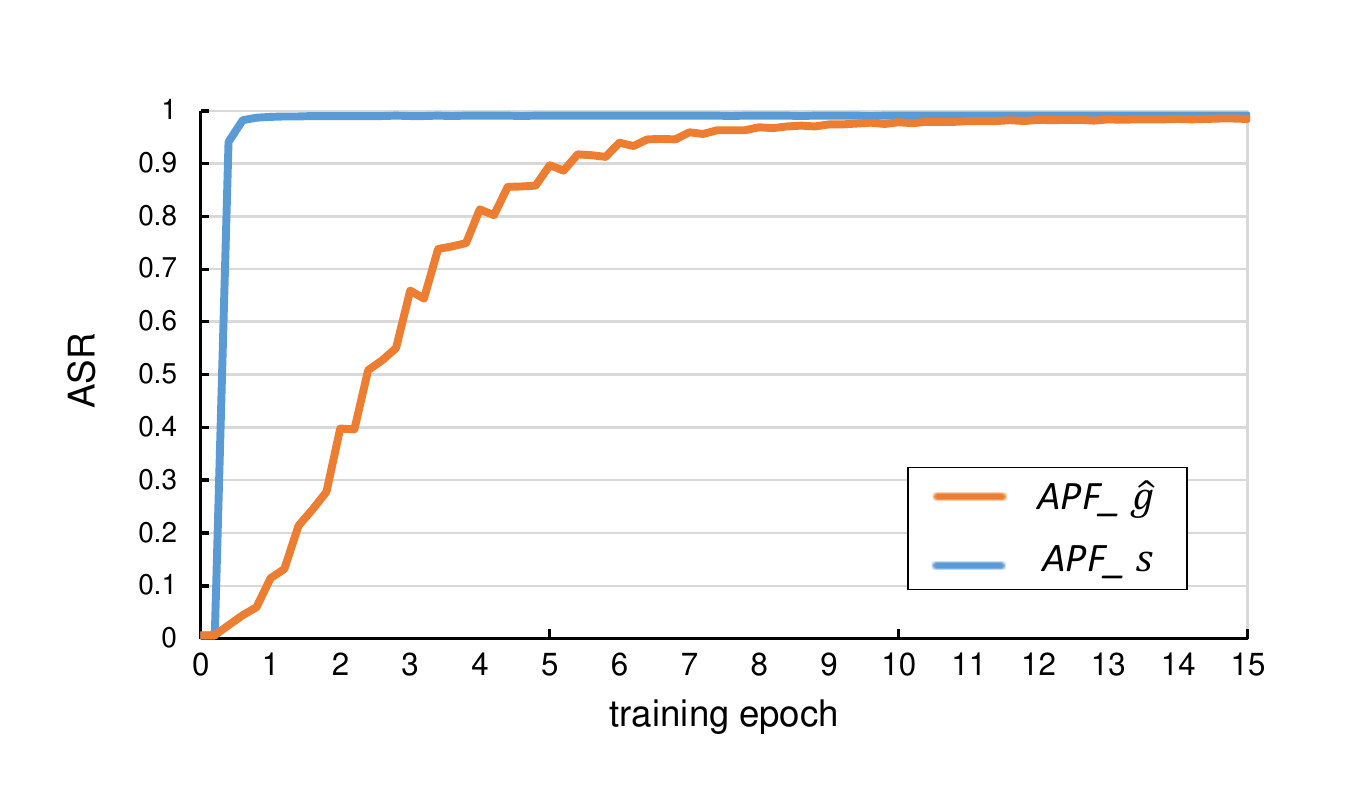}
  \end{center}
  \caption{Convergence curve w/ and w/o universal perturbation enhancement.}
  \label{fig:5}
\end{figure}

\subsubsection{The Influence of Universal Adversarial Perturbation Enhancement} \label{sec:4-2-2}
As mentioned in Section~\ref{sec:4-2-1}, universal perturbation can enhance the performance of adversarial images. In this subsection, we report further experimental results to examine the influence of universal perturbation enhancement. Testing ASRs of $APF\_{\hat{g}}$ and $APF\_s$ based on DI$^2$-FGSM of LFW dataset are shown in Fig.~\ref{fig:5} with respect to the training epochs. We can observe that: (1) The ASR curve of $APF\_s$ is almost always above $APF\_{\hat{g}}$, which is consistent with the observation from Table~\ref{tab:1} that $APF\_s$ outperforms $APF\_{\hat{g}}$. (2) $APF\_s$ converges nearly within an epoch, which is much faster than $APF\_{\hat{g}}$. To understand the mechanism behind, we analyze the training process of the network. For ease of description, the loss functions in Eqn.~\eqref{eq10} are defined as $L_s$. Universal adversarial perturbation $u$ will be transmitted to $L_s$ by forward propagation, then $\nabla_{\theta_T} L_s$ will affect the training of gradient transfer module by backward propagation. The enhancement module itself is easy to train due to very few parameters it has. So the convergence speed is accelerated. 

Moreover, three kinds of adversarial images \emph{Images\_u} perturbed by only the universal adversarial attack methods (UAP, GAP, RHP) are also implemented for comparison. As shown in Table~\ref{tab:1}, we can observe that these universal adversarial attack methods fail to achieve a reasonable ASR. This demonstrates that, while universal adversarial perturbation has potential to simultaneously attack multiple images, it singly cannot guarantee a promised attack performance compared with traditional attack methods utilizing image-specific information.

We propose a strategy that combining universal adversarial perturbation with image-dependent adversarial gradient, and achieve a productive performance. We hope this study could draw attention to this combination on the other adversarial attack tasks, e.g., classification, object detection, semantic segmentation.

\subsubsection{On \textbf{Non-accessibility}}\label{sec:4-2-3} 

To examine the potential of the proposed two-stage attack solution to approximate the traditionally generated adversarial perturbation, we further compare APF with the adversarial attack setting violating the \emph{non-accessibility} requirement, noted as \emph{original image accessible} in Table~\ref{tab:2}. For this setting, the original images are assumed to be exposed to the server, where adversarial images are generated by 5 adversarial gradient extraction server models. As shown in Table~\ref{tab:2}, the proposed end-cloud collaborated privacy-preserving solution ($APF\_s$) obtains comparable performance with the setting accessible to original images. This demonstrates that, even without directly accessing the original images, it is possible to achieve remarkable attack performance by conducting auxiliary operations such as gradient transfer and universal perturbation enhancement introduced in this study. This result on one hand opens up possibility for alternative way of implementing adversarial attack, but on the other hand imposes the new challenge to adversarial defense in the future when original samples are not available.

\begin{table}[t]
\centering
\caption{Black-box ASRs on different training-testing pairs.}\label{tab:2}
\begin{tabular}{c|cccc} 
\toprule[1pt]
\multirow{2}{*}{Training Models} &\multicolumn{4}{c}{Testing Models}\\
\cline{2-5}
 &FaceNet &SphereFace &ArcFace &Average\\
\hline
raw &  83.0\% & 50.2\% & 92.5\%& 75.1\% \\
\midrule
FaceNet  & 95.2\% & 75.6\% & 98.3\% & 89.6\% \\
SphereFace & 91.5\%& 85.6\%& 96.7\% & 91.1\% \\
ArcFace  & 93.8\% &73.3\% & 98.8\% &88.5\% \\
\midrule 
Ensemble(F+S) &95.1\% &81.8\% &\textbf{98.6\%} &\textbf{91.7\%}\\
\bottomrule[1pt]
\end{tabular}
\end{table}

\subsection{Real-world Implementation Discussion}\label{sec:4-3}
\subsubsection{Transferability and Black-box Attack}\label{sec:4-3-1}

Since what face recognition models the cracker use is unknown and typical crackers may use multiple face recognition models, in practice the privacy-preserving effectiveness essentially depends on the solution’s generalization and transferability under black-box attack settings. To address this, we implemented $3$ server models of ArcFace, FaceNet and SphereFace, and evaluated the attack performance on \emph{APF\_s}. The adversarial gradient extraction algorithm is fixed as DI$^2$-FGSM, and the ASRs are calculated based on the LFW dataset. To construct the black-box attack setting, among the $3$ server models, FaceNet and SphereFace are selected as white-box models, and ArcFace is selected as the black-box model due to its wide utilization in public Face ID system\footnote{~\small{The black-box model is to simulate the possible cracking face recognition choices in real-world applications. Therefore, a widely employed model can better evaluate the privacy-preserving performance in practice.}}. In addition, we also examined the performance of the ensemble adversarial attack combining FaceNet (F) and SphereFace (S). 

Table~\ref{tab:2} shows the black-box ASR under different training-testing pairs. For example, the value of $98.6\%$ represents the ASR trained with ensembled white-box models and tested on ArcFace. Higher ASR value means superior resistance performance to malicious face recognition model and better transferability of the method. There are several observations: (1) The adversarial images (the last four lines) generated by any model outperform than these adversarial images (first row) generated by naive adversarial gradient extraction algorithms. This demonstrates our proposed framework can improve the transferability of adversarial images, without limitation with specific models. (2) The adversarial images generated with a special model perform well when they are tested by the corresponding model, but slightly worse on the other models. It is expected that the white-box attack has better performance than black-box attack does. (3) When the adversarial images are generated by ensemble models and are attacked by ArcFace model (as the black-box), the ASR ($98.6\%$) is higher than the other two black-box methods ($95.1\%$, $81.8\%$), even almost catches up with the white-box ASR ($98.8\%$). It verifies the transferability of our proposed method in employing ensemble training towards black-box attacking. (4) The average ASRs of ensemble training model is the highest among all training models. It is expected with more models implemented in ensemble training, the ASR performance towards arbitrary black-box attacking methods will be guaranteed. Referring to previous the study~\cite{liu2017delving}, it is reasonable to choose the model with the large structure differences to employ ensemble adversarial training. In practical applications, we can carefully select widely-used white-box models with typically different structures to improve the generalization and transferability to specific models.

\begin{table}[t]
\centering
\caption{ASRs for \emph{adversarial-original} and \emph{adversarial-adversarial} matching.}\label{tab:4}
\begin{tabular}{cc|ccc} 
\toprule[1pt]
&&   \emph{Adversarial-original}  &  \emph{Adversarial-adversarial} \\
\midrule
\multirow{2}{*}{DI$^2$}&w/o $\delta$   & 90.2\% & $\mathbf{40.6\%}$ \\
&w/ $\delta$  & 86.7\% & 96.4\% \\
\midrule
\midrule
\multirow{2}{*}{M-DI$^2$}&w/o $\delta$   & 98.3\% & $\mathbf{27.2\%}$ \\
&w/ $\delta$  & 98.4\% & 94.7\% \\
\bottomrule[1pt]
\end{tabular}
\end{table}

\subsubsection{Adversarial-adversarial Image Matching}\label{sec:4-5}

The above experiments all assume that the enrolled image is original image without adversarial perturbation. However, in practice, there is a chance when the enrolled image collected by face crackers is already adversarially filtered, i.e., the problem turns to match between the adversarial images from unique user. To examine the performance of the proposed solution under this situation, we randomly selected $5,000$ testing images from $100$ subjects in the MS-Celeb-1M. Among the $50$ images for each subject, we set one image as the enrolled image and the remaining $49$ to generate the adversarial images.

Following Eqn.~\eqref{eq1} and~\eqref{eq2} to generate adversarial images, we evaluate the following two settings for each subject: (1) \emph{adversarial-original}, matching between the $49$ adversarial images and the original enrolled image; (2) \emph{adversarial-adversarial}, matching among the $49$ adversarial images. The resultant average ASR is reported in Table~\ref{tab:4} using adversarial attack methods of DI$^2$-FGSM and M-DI$^2$-FGSM (first row for each method). The results show that it is easy to match between two adversarial images for the same subject (with ASR of $40.6\%$ and $27.2\%$) and the proposed solution fails to attack the face cracker under this situation. We owe this result to that with the fixed feature of enrolled image $f_\theta(\mathbf{x_{e}})$ in Eqn.~\eqref{eq2}, the generated adversarial images from the $49$ original images tend to be similar and are easy to be grouped into one subject. Therefore, we introduce a random noise $\delta$ to modify Eqn.~\eqref{eq2} as follows:
\begin{equation}\label{eq14}
  L(\mathbf{x}, \mathbf{x_{e}}; \theta) = -d (f_\theta(\mathbf{x}), f_\theta(\mathbf{x_{e}})+\delta)
\end{equation}
where $\delta$ is a random vector with $10\%$ elements following a uniform distribution $U(-0.07, 0.07)$ and the remaining $90\%$ setting as $0$. $\delta$ acts as a random rotation deviation from the feature of the enrolled image, so to prevent the generated adversarial images overfitting along unique gradient direction. The results of generating adversarial examples via Eqn.~\eqref{eq14} are shown in the second row for each attack method in Table~\ref{tab:4}. It is observed that the introduction of $\delta$ significantly improves ASR for \emph{adversarial-adversarial} while basically maintains the high ASR for \emph{adversarial-original}. This modification guarantees the privacy-preserving effectiveness of our solution.

\section{Conclusion}

In this study, we introduce a portrait photo privacy-preserving solution when sharing to OSNs to resist malicious crawler and face recognition usage. The proposed end-cloud collaborated adversarial attack solution is validated to well satisfy all three requirements of \emph{privacy}, \emph{utility} and \emph{non-accessibility}. 

In the future, in addition to testing the solution’s effectiveness in practical trials, we are also interested to work towards the following two directions: (1) the combination of image-specific and image-independent perturbation in more attack scenarios, (2) the attack and defense attempts when the original sample is not accessible. 

\begin{acks}
This work is supported by the National Key R\&D Program of China (Grant No. 2018AAA0100604), and the National Natural Science Foundation of China (Grant No. 61632004, 61832002, 61672518, U19B2039, 61632006, 61772048, 61672071, and U1811463), and the Beijing Talents Project (2017A24), and the Beijing Outstanding Young Scientists Projects (BJJWZYJH01201910005018).
\end{acks}

\balance{
\bibliographystyle{ACM-Reference-Format}
\bibliography{sample-base}
}

\end{document}